\documentclass[reqno]{article}
\usepackage{amssymb,amsmath,cite,delarray,epsfig,hhline}
\begin{document}
\begin{flushright}
\begin{tabular}{l}
UMD-PP-03-010\\SLAC-PUB-9470
\end{tabular}
\end{flushright}
\begin{center}
{\LARGE \bf Leptogenesis And Neutrino Oscillations
Within A Predictive G(224)/SO(10)-Framework}
\vskip.25in
{\large{Jogesh C. Pati,\\ Department of Physics, University of Maryland,
College Park  MD 20742, USA\footnote{present address},\\
and\\ Stanford Linear Accelerator Center, Stanford University, Menlo Park, CA,
94025, USA.\\
August 15, 2002.}}
\end{center}

\vskip.25in
\begin{abstract}
A framework based on an effective symmetry that is either G(224)= 
SU(2$)_L\times$SU(2$)_R\times$SU(4$)^c$ or SO(10) has been proposed 
(a few years ago) that successfully describes
the masses and mixings of all fermions including neutrinos, with seven
predictions, in good accord with the data. Baryogenesis
via leptogenesis is considered within this framework by allowing for 
natural phases ($\sim 1/20$-1/2) in the entries of the Dirac and
Majorana mass-matrices. It is shown that the framework leads quite naturally, 
for both thermal as well as non-thermal leptogenesis, 
to the desired magnitude for the baryon asymmetry. This result is obtained in full accord with the
observed features of the atmospheric and solar neutrino oscillations, as well as with those of the quark and charged lepton masses and mixings, and the
gravitino-constraint. Hereby one
obtains a {\it unified description} of fermion masses, neutrino oscillations
and baryogenesis (via leptogenesis) within a single predictive framework.
\end{abstract}
%----------------------------------------------INTRODUCTION--------------
\newpage
{\large

\section{Introduction}
The observed matter-antimatter asymmetry of the universe
\cite{expr1,expr2} is an important clue
to physics at truly short distances. A natural understanding of its magnitude
(not to mention its sign) is thus a worthy challenge. Since the discovery of
the electroweak sphaleron effect \cite{Ruzmin}, baryogenesis via leptogenesis
\cite{Yanagida,Others} appears to be the most attractive and promising
mechanism to generate such an asymmetry \cite{GUT-baryo}. In the context of a unified theory
of quarks and leptons, leptogenesis involving decays of heavy right-handed
(RH)
neutrinos, is naturally linked to the masses of quarks and leptons, neutrino
oscillations and, of course, CP violation.

In this regard, the route to higher unification based on an effective
four-dimensional gauge symmetry of either
G(224)=SU(2)$_L\times$SU(2)$_R\times$SU(4)$^C$
\cite{PS}, or SO(10) \cite{Georgi} (that may emerge from a string
theory near the string scale and breaks spontaneously to the standard model
symmetry near the GUT scale \cite{FN1}) offers some distinct advantages, which
are directly relevant to understanding neutrino masses and implementing leptogenesis. These in particular include: (a) the
existence of the RH neutrinos as a compelling feature, (b) B-L as a local
symmetry, and (c) quark-lepton unification through SU(4)-Color. These three
features, first introduced in Ref. \cite{PS} in the context of the symmetry
G(224), are of course available within any symmetry that contains G(224) as
a subgroup; thus, they are available within SO(10) and E$_6$ \cite{Gursey},
though not in SU(5) \cite{GeorgiGlashow}. Effective symmetries such as
flipped SU(5)$\times$U(1) \cite{flipSU5} or [SU(3)]$^3$ \cite{SU3}, or
SU(2)$_{\rm L}\times$SU(2)$_{\rm R}\times$U(1)$_{\rm B\mbox{-}L}
\times$SU(3)$^{\rm C}$ \cite{Pati_etal} possess the first two features (a)
and (b), but not (c). Now, the {\it combination} of the four ingredients --
that is (i) the existence of the RH neutrino as an integral member of each family, (ii) the supersymmetric unification scale $M_X \sim 2\times 10^{16}\mbox{ GeV}$ \cite{SUSYunifscale} (which
provides the Majorana mass of the RH neutrinos), (iii) the symmetry
SU(4)-color (which provides the Dirac mass of the tau neutrino in terms of
the top quark mass), {\it and} (iv) the seesaw mechanism \cite{seesaw} -- yields
even quantitatively \cite{SeeJCP} just about the right value of
$\Delta m^2(\nu_2$-$\nu_3)$, as observed at SuperKamiokande
\cite{Superk}.

Furthermore, these three features (a)-(c) noted above also provide just the
needed ingredients -
that is superheavy $\nu_R$'s and spontaneous violation of B-L at high
temperatures - for implementing baryogenesis via leptogenesis.

Now, in a theory with RH neutrinos having heavy Majorana masses, the magnitude
of the lepton-asymmetry is known to depend crucially on both the Dirac as well
as Majorana mass matrices of the neutrinos \cite{SeeReview}. In this regard,
 a predictive G(224)/SO(10) framework, describing the masses and mixings of
all fermions, including neutrinos, has been proposed \cite{BPW} that appears
to be remarkably successful.
In particular it makes seven predictions including: $m_b(m_b)\approx 4.7-4.9$ GeV,
$m(\nu_3)\sim (1/20)$ eV(1/2-2), $V_{cb}\approx 0.044$,
$\sin^22\theta_{\nu_2\nu_3}^{\rm osc}\approx 0.9$-0.99, 
$V_{us}\approx 0.20$, $V_{ub}\approx 0.003$
 and $m_d\approx 8$ MeV, all 
in good accord with observations, to within 10\% (see Sec. 2). It
has been noted recently \cite{JCP} that the large angle MSW solution (LMA),
which is preferred by experiments \cite{LMA}, can arise quite plausibly
within
the same framework through SO(10)-invariant higher dimensional operators
which can contribute directly to the Majorana masses of the left-handed
neutrinos (especially to the $\nu_L^e\nu_L^\mu$ mixing mass) without
involving the familiar seesaw. 

As an additional point, it has been noted by Babu and myself
\cite{BabuJCPCascais} that the framework proposed in Ref. \cite{BPW} can
 naturally accomodate CP violation by introducing complex phases in the 
entries
of the fermion mass-matrices, which preserve the pattern of the mass-matrices
suggested in Ref. \cite{BPW} as well as its successes.

The purpose of the present paper is to estimate the lepton and thereby the
baryon excess that would typically be expected within this realistic
G(224)/SO(10)-framework for fermion masses and mixings
\cite{BPW,BabuJCPCascais}, by allowing for natural CP violating phases
($\sim$ 1/20 to 1/2, say) in the entries of the mass-matrices as in Ref.
\cite{BabuJCPCascais}. The goal would thus be to obtain a {\it unified
 description} of (a) fermion masses, (b) neutrino oscillations, and (c)
 leptogenesis within a single predictive framework \cite{FN2}.

 It should be noted that there have in fact been several attempts in the
 literature \cite{Incompletelist} at estimating the lepton and baryon
 asymmetries, many of which have actually been carried out in the context
 of SO(10) \cite{SO(10)}, though (to my knowledge) without an accompanying
 realistic framework for the masses and mixing of quarks, charged leptons as
 well as neutrinos \cite{FN3}. Also the results in these attempts as
 regards leptogenesis have not been uniformly encouraging \cite{23}.

 The purpose of this letter is to note that the G(224)/SO(10) framework, 
proposed in Ref. \cite{BPW} and \cite{BabuJCPCascais},
 leads quite naturally, for both thermal as well as non-thermal leptogenesis, to the desired magnitude for baryon asymmetry. This result is obtained in full accord with the 
observed features of atmospheric and solar neutrino oscillations,
 as well as with those of quark and charged lepton masses and mixings, and the gravitino-constraint. To
 present the analysis it would be useful to recall the salient features of
these prior works \cite{BPW,BabuJCPCascais} on fermion masses and mixings.
This is what is done in the next section.

\section{Fermion Masses and Neutrino Oscillations in G(224)/SO(10): A Brief
 Review of Prior Work}

The $3\times 3$ Dirac mass matrices for the four sectors $(u,d,l,\nu)$
proposed in Ref. \cite{BPW} were motivated in part by the notion that flavor
symmetries \cite{Hall} are responsible for the hierarchy among the elements
of these matrices (i.e., for "33"$\gg$"23"$\gg$"22"$\gg$"12"$\gg$"11", etc.),
and in part by the group theory of SO(10)/G(224), relevant to a minimal Higgs
system (see below). Up to minor variants \cite{FN4}, they are as follows:
\begin{eqnarray}
\label{eq:mat}
\begin{array}{cc}
M_u=\left[
\begin{array}{ccc}
0&\epsilon'&0\\-\epsilon'&\zeta_{22}^u&\sigma+\epsilon\\0&\sigma-\epsilon&1
\end{array}\right]{\cal M}_u^0;&
M_d=\left[
\begin{array}{ccc}
0&\eta'+\epsilon'&0\\ \eta'-\epsilon'&\zeta_{22}^d&\eta+\epsilon\\0&
\eta-\epsilon&1
\end{array}\right]{\cal M}_d^0\\
&\\
M_\nu^D=\left[
\begin{array}{ccc}
0&-3\epsilon'&0\\-3\epsilon'&\zeta_{22}^u&\sigma-3\epsilon\\
0&\sigma+3\epsilon&1\end{array}\right]{\cal M}_u^0;&
M_l=\left[
\begin{array}{ccc}
0&\eta'-3\epsilon'&0\\ \eta'+3\epsilon'&\zeta_{22}^d&\eta-3\epsilon\\0&
\eta+3\epsilon&1
\end{array}\right]{\cal M}_d^0\\
\end{array}
\end{eqnarray}
These matrices are defined in the gauge basis and are
multiplied by $\bar\Psi_L$ on left and $\Psi_R$ on right.
For instance, the row and column indices of $M_u$ are given by $(\bar u_L,
\bar c_L, \bar t_L)$ and $(u_R, c_R, t_R)$ respectively.
Note the group-theoretic up-down and quark-lepton correlations: the same 
$\sigma$ occurs in $M_u$ and $M_\nu^D$, and the same $\eta$ occurs in $M_d$
and $M_l$. It will become clear that the $\epsilon$ and $\epsilon'$ entries
are proportional to B-L and are antisymmetric in the family space 
(as shown above).
Thus, the same $\epsilon$ and $\epsilon'$ occur in both ($M_u$ and $M_d$) and
also in ($M_\nu^D$ and $M_l$), but $\epsilon\rightarrow -3\epsilon$ and
$\epsilon'\rightarrow -3\epsilon'$ as $q\rightarrow l$. Such correlations 
result in enormous reduction of parameters and thus in increased predictivity.
Such a patern for the mass-matrices can be obtained, using a minimal Higgs
system ${\bf 45}_H,{\bf 16}_H,{\bf \bar {16}}_H \mbox{ and }{\bf 10}_H $
 and a singlet S of SO(10), through effective couplings as follows \cite{FN26}:
\begin{multline}
\label{eq:Yuk}
{\cal L}_{\rm Yuk}=h_{33}{\bf 16}_3{\bf 16}_3{\bf 10}_H\\ +\left[
h_{23}{\bf 16}_2{\bf 16}_3{\bf 10}_H(S/M)+a_{23}{\bf 16}_2{\bf 16}_3{\bf 10}_H
({\bf 45}_H/M')(S/M)^p+g_{23}{\bf 16}_2{\bf 16}_3{\bf 16}_H^d
({\bf 16}_H/M'')(S/M)^q\right]\\+
\left[h_{22}{\bf 16}_2{\bf 16}_2{\bf 10}_H(S/M)^2+g_{22}{\bf 16}_2{\bf 16}_2
{\bf 16}_H^d({\bf 16}_H/M'')(S/M)^{q+1} \right]\\+
\left[g_{12}{\bf 16}_1{\bf 16}_2
{\bf 16}_H^d({\bf 16}_H/M'')(S/M)^{q+2}+
a_{12}{\bf 16}_1{\bf 16}_2
{\bf 10}_H({\bf 45}_H/M')(S/M)^{p+2}
\right]
\end{multline}
Typically we expect $M'$, $M''$ and $M$ to be of order $M_{\rm string}$
\cite{FN6}. The VEV's of $\langle{\bf 45}_H\rangle$ (along B-L),
$\langle{\bf 16}_H\rangle=\langle{\bf\bar {16}}_H\rangle$ (along standard 
model
singlet sneutrino-like component) and of the SO(10)-singlet $\langle S \rangle$
are of the GUT-scale, while those of ${\bf 10}_H$ and of the down type 
SU(2)$_L$-doublet component in ${\bf 16}_H$ (denoted by ${\bf 16}_H^d$) are
of the electroweak scale \cite{BPW,FN7}. Depending upon whether 
$M'(M'')\sim M_{\rm GUT}$ or $M_{\rm string}$ (see footnote \cite{FN6}), 
the 
exponent $p(q)$ is either one or zero \cite{FN8}.

The entries 1 and $\sigma$ arise respectively from $h_{33}$ and $h_{23}$
couplings, while $\hat\eta\equiv\eta-\sigma$ and $\eta'$ arise respectively
from $g_{23}$ and $g_{12}$-couplings. The (B-L)-dependent antisymmetric 
entries $\epsilon$ and $\epsilon'$ arise respectively from the $a_{23}$ and
$a_{12}$ couplings. [Effectively, with $\langle{\bf 45}_H\rangle\propto$ B-L,
the product ${\bf 10}_H\times{\bf 45}_H$ contributes as a {\bf 120}, whose 
coupling
is family-antisymmetric.] The small entry $\zeta_{22}^u$ arises from the 
$h_{22}$-coupling, while $\zeta_{22}^d$ arises from the joint contributions of
$h_{22}$ and $g_{22}$-couplings. As discussed in \cite{BPW}, using some of the 
observed masses as inputs, one obtains 
$|\hat\eta|\sim|\sigma|\sim|\epsilon|\sim {\cal O}(1/10)$, 
$|\eta'|\approx 4\times 10^{-3}$ and $|\epsilon'|\sim 2\times 10^{-4}$. The 
success of the framework presented in Ref. \cite{BPW} 
(which set $\zeta_{22}^u=\zeta_{22}^d=0$) in describing fermion masses and
mixings remains essentially unaltered if 
$|(\zeta_{22}^u,\zeta_{22}^d)|\leq (1/3)(10^{-2})$ (say). 

Such a hierarchical form of the mass-matrices, with $h_{33}$-term being
dominant, is attributed in part to flavor gauge symmetry(ies) that
distinguishes between the three families \cite{FN9}, and in part to higher
dimensional operators involving for example $\langle{\bf 45}_H\rangle/M'$
or $\langle{\bf 16}_H\rangle/M''$, which are supressed by 
$M_{\rm GUT}/M_{\rm string}\sim 1/10$, if $M'$ and/or 
$M''\sim M_{\rm string}$. 

To discuss the neutrino sector one must specify the Majorana mass-matrix of
the RH neutrinos as well. These arise from the effective couplings of the
form
\cite{FN30}:
\begin{eqnarray}
\label{eq:LMaj}
{\cal L}_{\rm Maj}=f_{ij}{\bf 16}_i{\bf 16}_j{\bf\bar{16}}_H{\bf\bar{16}}_H/M
\end{eqnarray}
where the $f_{ij}$'s include appropriate powers of $\langle S \rangle/M$, in
accord with flavor charge assignments of ${\bf 16}_i$ (see \cite{FN9}). For
the $f_{33}$-term to be leading, we must assign the charge $-a$ to 
${\bf\bar{16}}_H$. This leads to a hierarchical form for the Majorana
mass-matrix
\cite{BPW}:
\begin{eqnarray}
\label{eq:MajMM}
M_R^\nu=\left[
\begin{array}{ccc}
x&0&z\\0&0&y\\z&y&1
\end{array}
\right]M_R
\end{eqnarray}
Following the flavor-charge assignments given in footnote \cite{FN9}, we 
expect $|y|\sim
\langle S/M\rangle\sim 1/10$, 
$|z|\sim (\langle S/M\rangle)^2\sim$ (1/200)(1 to 
1/2, say),  $|x|\sim (\langle S/M\rangle)^4\sim (10^{-4}$-$10^{-5})$ (say).
The "22" element (not shown) is $\sim (\langle S/M\rangle)^2$ and its
magnitude is
taken to be $< |y^2/3|$, while the "12" element (not shown) is
$\sim (\langle S/M\rangle)^3$.. We expect 
\begin{eqnarray} \label{MR}
M_R=f_{33}\langle {\bf \bar{16}}_H \rangle^2/M_{\rm string}\approx (10^{15}\mbox{ GeV})(1/2-2) 
\end{eqnarray}
for $\langle{\bf \bar{16}}_H\rangle\approx 2\times 10^{16}$ GeV, 
$M_{\rm string}\approx 4\times 10^{17}$ GeV \cite{Kaplunovsky} and 
$f_{33}\approx 1$. Allowing for 2-3 mixing, this value of $M_R$ [together with
the SU(4)-color relation $m(\nu_3^{\rm Dirac})=m_t(M_{\rm GUT})\approx 120$
 GeV] 
leads to $m(\nu_3)\approx (1/24$ eV)(1/2-2) \cite{BPW,SeeJCP,JCPErice}, in good 
accord with the SuperK data.

Ignoring possible phases in the parameters and thus the source of CP violation
for a
moment, as was done in Ref. \cite{BPW}, the parameters $(\sigma,\eta,
\epsilon, \epsilon',\eta', {\cal M}_u^0, {\cal M}_D^0,\mbox{ and } y)$ can be 
determined
by using, for example, $m_t^{\rm phys}=174$ GeV, $m_c(m_c)=1.37$ GeV,
$m_S(1\mbox{ GeV})=110$-116 MeV, $m_u(1\mbox{ GeV})=6$ MeV, the observed
masses of $e$, $\mu$, and $\tau$ and
$m(\nu_2)/m(\nu_3)\approx 1/(7\pm1)$ (as suggested by a combination
of atmospheric and solar neutrino data, the latter corresponding to the LMA MSW solution, see
below) as inputs. One is thus led, {\it for this CP conserving case}, to the
 following fit for the parameters, and the
associated predictions \cite{BPW}. [In this fit, we drop 
$|\zeta_{22}^{u,d}|\lesssim (1/3)(10^{-2})$ and leave the small quatities 
$x$ and
$z$ in $M_R^\nu$ undetermined and proceed by assuming that they have the
magnitudes suggested by flavor symmetries
(i.e., $x\sim (10^{-4}$-$10^{-5})$ and $z\sim$ (1/200)(1 to 1/2)
(see remarks below Eq. \eqref{eq:MajMM})]:
\begin{eqnarray}
\label{eq:fit}
\begin{array}{c}
\sigma\approx 0.110, \quad \eta\approx 0.151, \quad \epsilon\approx -0.095,
 \quad |\eta'|\approx 4.4 \times 10^{-3},\\
\begin{array}{cc}
\epsilon'\approx 2\times 10^{-4},& {\cal M}_u^0\approx m_t(M_X)\approx 120
\mbox{ GeV}, \\{\cal M}^0_D\approx m_b(M_X)\approx 1.5 \mbox{ GeV}, & 
y\approx -1/17.
\end{array}
\end{array}
\end{eqnarray}
These in turn lead to the following predictions for the quarks and
light neutrinos \cite{BPW, JCPErice}:
\begin{eqnarray}
\label{eq:pred}
\begin{array}{l}
m_b(m_b)\approx(4.7\mbox{-}4.9)\mbox{ GeV},\\
\sqrt{\Delta m_{23}^2} \approx m(\nu_3)\approx\mbox{(1/24 eV)(1/2-2)},\\
V_{cb}\approx\left|\sqrt{\frac{m_s}{m_b}}\left|\frac{\eta+\epsilon}
{\eta-\epsilon}\right|^{1/2}- \sqrt{\frac{m_c}{m_t}}\left|\frac{\sigma
+\epsilon}{\sigma-\epsilon}\right|^{1/2}\right|\approx 0.044,\\
\left\{ \begin{array}{l}
\theta^{\rm osc}_{\nu_{\mu}\nu_{\tau}}\approx\left|\sqrt{\frac{m_\mu}{m_\tau}}
\left|\frac{\eta-3\epsilon}{\eta+3\epsilon}\right|^{1/2}+
\sqrt{\frac{m_{\nu_2}}{m_{\nu_3}}}\right|\approx |0.437+(0.378\pm 0.03)|,\\
\mbox{Thus, } \sin^2 2\theta^{\rm osc}_{\nu_{\mu}\nu_{\tau}}\approx 0.99, \mbox{   (for $m(\nu_2)/m(\nu_3)\approx 1/7$),}\\
\end{array}\right.\\
V_{us}\approx \left|\sqrt{\frac{m_d}{m_s}}-\sqrt{\frac{m_u}{m_c}}\right|
\approx 0.20,\\
\left|\frac{V_{ub}}{V_{cb}} \right|\approx \sqrt{\frac{m_u}{m_c}}\approx
0.07,\\
m_d(\mbox{1 GeV})\approx \mbox{8 MeV},\\
\theta^{\rm osc}_{\nu_e\nu_{\mu}}\approx 0.06
\mbox{  (ignoring non-seesaw contributions); see remarks below.}
\end{array}
\end{eqnarray}

The Majorana masses of the RH neutrinos ($N_{iR}\equiv N_i$) are given by \cite{JCPErice}:
\begin{eqnarray}
\label{eq:MajM}
M_{3}& \approx & M_R\approx 10^{15}\mbox{ GeV (1/2-1)},\nonumber\\
M_{2}& \approx & |y^2|M_{3}\approx \mbox{(2.5$\times 10^{12}$ GeV)(1/2-1)},\\
M_{1}& \approx & |x-z^2|M_{3}\sim (1/2\mbox{-}2)10^{-5}M_{3}\sim
\mbox{$10^{10}$ GeV(1/4-2)}.\nonumber
\end{eqnarray}

Note that we necessarily have a hierarchical pattern for the light as well as the heavy neutrinos (see discussions below on $m_{\nu_1}$). Leaving out the $\nu_e$-$\nu_2$ oscillation angle for a moment, it seems
remarkable that the first seven predictions in Eq. \eqref{eq:pred} agree
with observations, to within 10\%. Particularly intriguing is the (B-L)-dependent 
{\it group-theoretic correlation} between the contribution from the first
term in $V_{cb}$ and that in $\theta^{\rm osc}_{\nu_2\nu_3}$, which
explains simultaneously why one is small ($V_{cb}$) and the other is
large ($\theta^{\rm osc}_{\nu_2\nu_3}$) \cite{newFN36}. That in turn
provides some degree of confidence in the gross structure of the
mass-matrices.

As regards $\nu_e$-$\nu_{\mu}$ and $\nu_e$-$\nu_{\tau}$ oscillations, the 
standard seesaw mechanism would typically lead to rather small angles as in  
Eq.~\eqref{eq:pred}, within the framework presented above \cite{BPW}.
It has, however, been noted recently \cite{JCP} that small intrinsic 
(non-seesaw) masses $\sim 10^{-3}$ eV of the LH neutrinos can arise quite
plausibly through higher dimensional operators of the form \cite{FN32}:
\(W_{12}\supset \kappa_{12}{\bf 16}_1{\bf 16}_2{\bf 16}_H{\bf 16}_H{\bf 10}_H
{\bf 10}_H/M_{\rm eff}^3\), 
without involving the standard seesaw mechanism \cite{seesaw}.
One can verify that such a term would lead to an intrinsic Majorana mixing
mass term of the form $m_{12}^0\nu_L^e\nu_L^\mu$, with a strength given by
\(m_{12}^0\approx \kappa_{12}\langle{\bf 16}_H \rangle^2(175\mbox{ GeV})^2/
M_{\rm eff}^3\sim (1.5\mbox{-}6)\times 10^{-3}\) eV, for 
$\langle{\bf 16}_H \rangle\approx (1\mbox{-}2)M_{\rm GUT}$ and 
\(\kappa_{12}\sim 1$, if $M_{\rm eff}\sim M_{\rm GUT}\approx 2\times 10^{16}\)
GeV \cite{FN33}. Such an intrinsic Majorana $\nu_e\nu_{\mu}$ mixing mass
$\sim $ few$\times 10^{-3}$ eV, though small compared to $m(\nu_3)$, is
still much larger than what one would generically get for the corresponding
term from the standard seesaw mechanism [as in Ref.~\cite{BPW}]. Now, the
diagonal ($\nu_{\mu}\nu_{\mu}$) mass-term, arising from standard seesaw can 
naturally be $\sim$ (3-8)$\times 10^{-3}$ eV for $|y|\approx 1/20$-1/15, say
\cite{BPW}. Thus, taking the net values of 
\(m_{22}^0\approx (6-7)\times 10^{-3}\) eV,  
$m_{12}^0\sim 3\times 10^{-3}$ eV  as above and 
$m_{11}^0\ll 10^{-3}$ eV, which are all plausible, we obtain 
\(m_{\nu_2}\approx(6-7)\times 10^{-3}\) eV, 
\(m_{\nu_1}\sim \mbox{(1 to few)} \times 10^{-3}\) eV, so that 
\(\Delta m^2_{12}\approx (3.6\mbox{-}5)\times 10^{-5} \mbox{ eV}^{2}\) and 
\(\sin^{2} 2\theta_{\nu_{e}\nu{\mu}}^{\rm osc}\approx 0.6-0.7\). These go
well with the LMA MSW solution of the solar neutrino puzzle.

In summary, {\it the intrinsic non-seesaw contribution} to the Majorana
masses of the
LH neutrinos can possibly have the right magnitude for $\nu_e$-$\nu_{\mu}$
mixing so as to lead to the LMA solution within the G(224)/SO(10)-framework,
without upsetting the successes of the first seven predictions 
in Eq.~\eqref{eq:pred}. [In contrast to the near maximality 
of the 
$\nu_{\mu}$-$\nu_{\tau}$ oscillation angle, 
however, which emerges as a compelling
prediction of the framework \cite{BPW}, the LMA solution, as obtained above, 
should, be regarded as a consistent possibility, rather than as a compelling prediction, within this framework.]

Before discussing leptogenesis, we need to discuss the origin of CP 
violation within the G(224)/SO(10)-framework presented above. The discussion 
so far
has ignored, for the sake of simplicity, possible CP violating phases in the
parameters ($\sigma$, $\eta$, $\epsilon$, $\eta'$, $\epsilon'$,
$\zeta_{22}^{u,d}$, $y$, $z$, and $x$) of the Dirac and Majorana mass matrices
[Eqs. \eqref{eq:mat}, and \eqref{eq:MajMM}]. In general, however, these
parameters can and generically will have phases \cite{FN34}. Some combinations
of these phases enter into the CKM matrix and define the Wolfenstein parameters
$\rho_W$ and $\eta_W$ \cite{Wolfenstein}, which in turn induce CP violation by
utilizing the standard model interactions. As observed in Ref. 
\cite{BabuJCPCascais}, an additional and potentially important source of CP 
and flavor violations (as in $K^0\leftrightarrow\bar K^0$, 
$B_{d,s}\leftrightarrow\bar B_{d,s}$, $b\rightarrow s\bar s s$, etc. 
transitions)
arise in the model through supersymmetry \cite{FN36}, involving squark and
gluino loops (box and penguin), simply because of the embedding of MSSM within
a string-unified G(224) or SO(10)-theory near the GUT-scale, and the 
assumption that primordial SUSY-breaking occurs near the string scale
($M_{\rm string}>M_{\rm GUT}$) \cite{FN37}.
 It is shown
that complexification of the parameters ($\sigma$, $\eta$, $\epsilon$,
$\eta'$, $\epsilon'$, etc.), through introduction of phases $\sim 1/20$-1/2
(say) in them, can still preserve the successes of the predictions as regards
fermion masses and neutrino oscillations shown in Eq. \eqref{eq:pred}, as
long as one maintains nearly the magnitudes of the real parts of the
parameters and especially their relative signs as obtained in Ref. \cite{BPW}
and shown in Eq. \eqref{eq:fit} \cite{FN38}. Such a picture is also in accord
with the observed features of CP and flavor violations in $\epsilon_K$, 
$\Delta m_{Bd}$, and asymmetry parameter in
 $B_d\rightarrow J/\Psi+K_s$, while predicting observable new effects in
 processes such as $B_s\rightarrow \bar B_s$ and $B_d\rightarrow \Phi+K_s$
 \cite{BabuJCPCascais}.

 We therefore proceed to discuss leptogenesis concretely within the framework
 presented above by adopting the Dirac and Majorana fermion mass matrices
 as shown in Eqs. \eqref{eq:mat} and  \eqref{eq:MajMM} and assuming that the
 parameters appearing in these matrices can have natural phases
 $\sim 1/20$-1/2 (say) with either sign up to addition of $\pm \pi$, {\it
while their real
 parts have the relative signs and nearly the magnitudes given in Eq.
 \eqref{eq:pred}.}

\section{Leptogenesis}

In the context of an inflationary scenario \cite{d}, with a plausible reheat
temperature $T_{RH}\sim (1 \mbox{ to few})\times 10^{9}$ GeV (say), one can
avoid the
well known gravitino problem if $m_{3/2}\sim (1\mbox{ to }2)$ TeV 
\cite{gravitino} and
yet produce the lightest heavy neutrino $N_1$ efficiently from the thermal
bath if $M_{1}\sim\mbox{(3 to 5)}
\times 10^{9}$ GeV (say), in accord with Eq.~\eqref{eq:MajM}
[$N_2$ and $N_3$ are of course too heavy to be produced at $T\sim T_{RH}$].
Given lepton number (and B-L) violation occurring through the Majorana mass of
$N_1$, and C and CP violating phases in the Dirac and/or Majorana fermion
mass-matrices as mentioned above, the
out-of-equilibrium decays of $N_1$ (produced from the thermal bath) into
$l+H$ and $\bar l+\bar H$ and into the corresponding SUSY modes
$\tilde l+\tilde H$ and $\bar {\tilde l}+\bar {\tilde H}$ would produce
a B-L violating lepton asymmetry; so also would the decays of $\tilde N_1$
and $\bar{\tilde{N_1}}$. Part of this asymmetry would of course be washed
out due to inverse decays and lepton number violating
2$\leftrightarrow$2-scatterings. We will assume this commonly adopted
mechanism for the so-called thermal leptogenesis (At the end, we will,
however, consider an
interesting alternative that would involve non-thermal leptogenesis). This
mechanism has been extended to incorporate supersymmetry by several authors
(see e.g., \cite{Campbell,Covi,Plumacher}). The net lepton asymmetry of the
universe [$Y_L\equiv(n_L-n_{\bar L})/s$] arising from decays of $N_1$ into
 $l+H$ and $\bar l+\bar H$ and into the corresponding SUSY modes
($\tilde l+\tilde H$ and $\bar {\tilde l}+\bar {\tilde H}$) and likewise from
$(\tilde N_1, \bar{\tilde{N_1}})$-decays \cite{Campbell,Covi,Plumacher} is
given by:
\begin{eqnarray}
\label{eq:YL}
Y_L=\kappa\epsilon_1\left(\frac{n_{N_1}+n_{\tilde N_1}+
n_{\bar{\tilde{N_1}}}}{s}\right)\approx \kappa\epsilon_1/g^*
\end{eqnarray}
where $\epsilon_1$ is the lepton-asymmetry produced per $N_1$
(or $(\tilde N_1+\bar{\tilde{N_1}})$-pair) decay
(see below), $\kappa$ is an efficiency or damping factor that represents
the washout effects mentioned above (thus $\kappa$ incorporates the
extent of departure from thermal equilibrium in $N_1$-decays; such a departure is needed to realize lepton asymmetry), and $g^*\approx 228$ is
the number of light degrees of freedom in MSSM.

The lepton asymmetry $Y_L$ is converted to baryon asymmetry, by the sphaleron
effects, which is given by:
\begin{eqnarray}
\label{eq:YB}
Y_B=\frac{n_B-n_{\bar B}}{s}=C\,Y_L,
\end{eqnarray}
where, for MSSM, $C\approx -1/3$.
Taking into account the inteference between the tree and loop-diagrams for
the decays of $N_1\rightarrow lH$ and $\bar l\bar H$ (and likewise for
$N_1\rightarrow \tilde l\tilde H$ and $\bar{\tilde l}\bar{\tilde H}$ modes
and also for $\tilde N_1$ and $\bar{\tilde{N_1}}$-decays), the CP violating
lepton asymmetry parameter in each of the four channels (see e.g.,
\cite{Covi} and \cite{Plumacher}) is given by
\begin{eqnarray}
\label{eq:epsilon1}
\epsilon_1=\frac{1}{8\pi v^2(M_D^\dagger M_D)_{11}}\sum_{j=2,3}
{\rm Im} \left[(M_D^\dagger M_D)_{j1} \right]^2 f(M_j^2/M_1^2)
\end{eqnarray}
where $M_D$ is the Dirac neutrino mass matrix evaluated in a basis in which 
the Majorana mass matrix of the RH neutrinos 
$M_R^\nu$ [see Eq. \eqref{eq:MajMM}] is diagonal, $v=(\mbox{174 GeV})
\sin\beta$ and the function $f\approx -3(M_1/M_j)$ for the case of SUSY with
$M_j\gg M_1$.

The efficiency factor mentioned above, is often expressed in terms of the
parameter $K\equiv [\Gamma(N_1)/2H]_{T=M_1}$ \cite{d}. Assuming initial
thermal abundance for $N_1$, $\kappa$ is normalized so that it is 1 if
$N_1$'s decay fully out of equilibrium corresponding to $K\ll 1$
(in practise, this actually requires $K<0.1$). Including inverse decays as
well as $\Delta L\neq 0$-scatterings in the Boltzmann equations, a recent
analysis \cite{Bari} shows that in the relevant parameter-range of interest
to us (see below), the efficiency factor (for the SUSY case) is given by
\cite{newFN50}:
\begin{eqnarray}
\label{eq:d}
\kappa\approx (0.7\times 10^{-4})({\rm eV}/\tilde{m_1})
\end{eqnarray}
where $\tilde{m_1}$ is an effective mass parameter (related to $K$
\cite{newFN51}), and is given by \cite{m_tilde}:
\begin{eqnarray}
\label{eq:k}
\tilde{m_1}\equiv (m^\dagger_Dm_D)_{11}/M_1.
\end{eqnarray}
Eq. \eqref{eq:k} should hold to better than 20\% (say), when 
$\tilde m_1\gg 5\times 10^{-4}$ eV \cite{Bari} 
(This applies well to our case, see below).

Given the Dirac and Majorana mass matrices of the neutrinos [Eqs. 
\eqref{eq:mat} and \eqref{eq:MajMM}], we are now ready to evaluate lepton
assymetry by using Eqs. \eqref{eq:YL}-\eqref{eq:k}. 

The Majorana mass matrix [Eq. \eqref{eq:MajMM}] describing the mass-term 
$\nu^T_RCM_R^\nu\nu_R$ is diagonalized by the transformation 
$\nu_R=U_R^{(1)}U_R^{(2)}N_R$,
where (to a good approximation)
\begin{eqnarray}
\label{eq:UR}
U_R^{(1)}\approx\left[\begin{array}{ccc}1&0&z\\0&1&y\\-z&-y&1\end{array}
\right],
\end{eqnarray}
and $U_R^{(2)}={\rm diag}(e^{i\phi_1},e^{i\phi_2},e^{i\phi_3})$ is a diagonal
phase matrix that ensures real positive eigenvalues. The phases $\phi_i$ can
of course be derived from those of the parameters in $M_R^\nu$ [see Eq. 
\eqref{eq:MajMM}].
Applying this transformation to the neutrino Dirac mass-term 
$\bar\nu_LM_\nu^D\nu_R$ given by Eq. \eqref{eq:mat}, we obtain
$M_D=M_\nu^DU_R^{(1)}U_R^{(2)}$, which appears in Eqs.
\eqref{eq:epsilon1} and \eqref{eq:k}. In turn, this yields:
\begin{eqnarray}
\label{eq:MDMD21}
\frac{(M_D^\dagger M_D)_{21}}{\left({\cal M}_u^0\right)^2}&=&
e^{i(\phi_1-\phi_2)}\{
\left(-3{\epsilon'}^*
-\zeta^*_{13}y^*\right)\left(\zeta_{11}-z\zeta_{13}\right)
\nonumber\\&+&
\left[\zeta_{22}^{u*}-y^*\left(\sigma^*-3\epsilon*\right)\right]
\left[3\epsilon'-z\left(\sigma-3\epsilon\right)\right]+
\left(\zeta_{31}-z\right)\left[\left(\sigma^*+3\epsilon^*\right)-y^*\right]\}\\
\label{eq:MDMD11}
\frac{(M_D^\dagger M_D)_{11}}{\left({\cal M}_u^0\right)^2}&=&
\left|3\epsilon'-z(\sigma-3\epsilon)\right|^2+\left|\zeta_{31}-z\right|^2
\end{eqnarray}
In writing Eqs. \eqref{eq:MDMD21} and \eqref{eq:MDMD11}, we have allowed, for 
the sake of generality, the relatively small ``11'', ``13'', and ``31''
elements in the Dirac mass-matrix $M_\nu^D$, denoted by $\zeta_{11}$,
$\zeta_{13}$ and $\zeta_{31}$ respectively, which are not exibited in Eq.
\eqref{eq:mat}. Guided by considerations of flavor symmetry
(see footnote \cite{FN9}), we would expect 
$|\zeta_{11}|\sim(\langle S\rangle/M)^4\sim 10^{-4}$-$10^{-5}$, and
$|\zeta_{13}|\sim|\zeta_{31}|\sim(\langle S\rangle/M)^2\sim 10^{-2}$(1 to
1/3) (say). These small elements (neglected in \cite{BPW}) would not, 
of course, have any noticeable effects on the predictions of the fermion
masses and mixings given in Eq. \eqref{eq:pred}, except possibly on $m_d$.

We now proceed to make numerical estimates of lepton and baryon-asymmetries
by taking the magnitudes and the relative signs of the real parts of the
parameters ($\sigma$, $\eta$, $\epsilon$, $\eta'$, $\epsilon'$, and $y$)
approximately the same as in Eq. \eqref{eq:fit}, but allowing in general
for natural phases in them. As mentioned before [see for example the fit
given in footnote \cite{FN38} and Ref. \cite{BabuJCPCascais} (to appear)]
such a procedure introduces CP violation in accord with observation, while
preserving the successes of the framework as regards its predictions for
fermion masses and neutrino oscillations \cite{BabuJCPCascais, BPW}.

Given the magnitudes of the parameters (see Eqs. \eqref{eq:fit} and 
Ref. \cite{FN38}), which are obtained from considerations of fermion masses
and neutrino oscillations \cite{BPW,BabuJCPCascais} -- that is 
$|\sigma|\approx |\epsilon|\approx 0.1$, $|y|\approx 0.06$, 
$|\epsilon'|\approx 2\times 10^{-4}$, $|z|\sim (1/200)(1\mbox{ to }1/2)$,
$|\zeta_{22}^u|\sim 10^{-3}(1\mbox{ to }3)$, $|\zeta_{13}|\sim|\zeta_{31}|
\sim(1/200)(1\mbox{ to }1/2)$, with  the real parts of ($\sigma$, $\epsilon$ 
and $y$) having the signs (+, -, -) respectively, we would expect the typical
magnitudes of the three terms of Eq. \eqref{eq:MDMD21} to be as follows:
\begin{eqnarray}
\label{eq:terms21}
|\mbox{$1^{st}$ Term}|&=&\left|\left(-3{\epsilon'}^*-\zeta_{13}^*y^*\right)
\left(\zeta_{11}-z\zeta_{13}\right)\right|\nonumber\\
&\approx& \left[(6\mbox{ to }8)\times 10^{-4}\right]\left[(2.5\times 10^{-5})
(1\mbox{ to }1/4)\right]\sim 10^{-8}\nonumber\\
|\mbox{$2^{nd}$ Term}|&=&\left|\left\{\zeta_{22}^{u*}-y^*\left(\sigma^*-
3\epsilon^*\right)\right\} \left\{3\epsilon'-z(\sigma-3\epsilon)
\right\}\right|\\
&\approx& \left(2\times 10^{-2}\right)\left[2\times 10^{-3}(1\mbox{ to }1/2)
\right]\approx  \left(4\times 10^{-5}\right)(1\mbox{ to }1/2)\nonumber\\
|\mbox{$3^{rd}$ Term}|&=&\left|\left(\zeta_{31}-z\right)\left\{
\left(\sigma^*+3\epsilon^*\right)-y^*\right\}\right|\nonumber\\
&\approx& [(1/200)(1/2\mbox{ to }1/5)](0.13)\approx\left(0.7\times 10^{-3}
\right)(1/2\mbox{ to }1/5)\nonumber
\end{eqnarray}
Thus, assuming that the phases of the different terms are roughly comparable,
the third term would clearly dominate. The RHS of Eq. \eqref{eq:MDMD11} is 
similarly estimated to be:
\begin{eqnarray}
\label{eq:terms11}
\frac{\left( M_D^\dagger M_D\right)_{11}}{\left({\cal M}_u^0\right)^2}&=&
|3\epsilon'-z(\sigma-3\epsilon)|^2+\left|\zeta_{31}-z\right|^2\nonumber\\
&\approx &\left|6\times 10^{-4}\mp 2\times 10^{-3} (1\mbox{ to }1/2)\right|^2+
\left|5\times 10^{-3}(1/2\mbox{ to }1/5)\right|^2\\
&\approx & 2.5\times 10^{-5}(1/4\mbox{ to }1/6)\nonumber
\end{eqnarray}
Since $|\zeta_{31}|$ and $|z|$ are each expected to be of order 
(1/200)(1 to 1/2), we have allowed in Eqs. \eqref{eq:terms21} and
\eqref{eq:terms11} for a possible mild cancellation between
their contributions to $|\zeta_{31}-z|$ by putting
$|\zeta_{31}-z|\approx$ (1/200)(1/2 to 1/5) (say).
In going from the second to the third step of Eq. \eqref{eq:terms11} we have
assumed (for simplicity) that the second term of $(M_D^\dagger M_D)_{11}/
({\cal M}_u^0)^2$ given by $|\zeta_{31}-z|^2$ denominates over the first.
This in fact holds for a large part of the expected parameter space,
especially for values of $|z|\approx (1/200)(1/2)\lesssim |\zeta_{31}|
\approx (1/200)$(1 to 3/4) (say). Note that the combination $|\zeta_{31}-z|$
also enters into the dominant term [i.e., the third term in Eq.
\eqref{eq:terms21}] of  $(M_D^\dagger M_D)_{21}/({\cal M}_u^0)^2$. As a
result, to a good approximation (in the region of parameter space mentioned
above), the lepton-asymmetry parameter $\epsilon_1$ [given by Eq.
\eqref{eq:epsilon1}] becomes independent of the magnitude of
$|\zeta_{31}-z|^2$, and is given by:
\begin{eqnarray}
\label{eq:epsilon2}
\epsilon_1\approx \frac{1}{8\pi}\left(\frac{{\cal M}_u^0}{v}\right)^2
|(\sigma+3\epsilon)-y|^2\sin\left(2\phi_{21}\right)(-3)
\left(\frac{M_1}{M_2}\right)\approx -\left(2.0\times 10^{-6}\right)
\sin\left(2\phi_{21}\right),
\end{eqnarray}
where, $\phi_{21}={\rm arg}[(\zeta_{31}-z)(\sigma^*+3\epsilon^*-y^*)]+
(\phi_1-\phi_2)$, and
we have put $({\cal M}_u^0/v)^2\approx 1/2$, $|\sigma+3\epsilon-y|\approx
0.13$ (see Eq. \eqref{eq:fit} and Ref. \cite{FN38}), and
for concreteness (for the present case of thermal leptogenesis)
$M_1\approx 4\times 10^9\mbox{ GeV}$ and $M_2\approx 2\times 10^{12}
\mbox{ GeV}$ [see Eq. \eqref{eq:MajM}]. The parameter
$\tilde{m_1}$, given by Eq. \eqref{eq:k}, is (approximately) proportional to
$|\zeta_{31}-z|^2$ [see Eq. \eqref{eq:terms11}].
It is given by:
\begin{eqnarray}
\label{eq:k_}
\tilde{m_1}\approx |\zeta_{31}-z|^2 ({\cal M}_u^0)^2/M_1\approx
(1.9\times 10^{-2}\mbox{ eV})(\mbox{1 to 1/6}) \bigg(\frac{4\times 10^9 \mbox{ GeV}}{M_1}\bigg)
\end{eqnarray}
where, as before, we have put $|\zeta_{31}-z|\approx (1/200)(1/2\mbox{ to }1/5)$. The corresponding
efficiency
factor $\kappa$ [given by Eq. \eqref{eq:d}], lepton and baryon-asymmetries
$Y_L$ and $Y_B$ [given by Eqs. \eqref{eq:YL} and \eqref{eq:YB}] and the 
requirement on the phase-parameter $\phi_{21}$ are listed in Table~\ref{tab:phase}.
\begin{table}
\begin{center}
\begin{tabular}{|c|c|c|c|c|}
\hline
&\multicolumn{3}{c|}{$|\zeta_{31}-z|$}\\
\hhline{|~|-|-|-|}
&(1/200)(1/3)&(1/200)(1/4)&(1/200)(1/5)\\ \hline\hline
$\tilde{m}_{1}$(eV) & $0.83\times 10^{-2}$ & $0.47\times 10^{-2}$ &
$0.30\times 10^{-2}$ \\ \hline
$\kappa$ & 1/73 & 1/39&1/24 \\ \hline
$Y_L/\sin(2\phi_{21})$ & $-11.8\times 10^{-11}$ & $-22.4\times
10^{-11}$ & $-36\times 10^{-11}$ \\ \hline
$Y_B/\sin(2\phi_{21})$ & $4\times 10^{-11}$ & $7.5\times 10^{-11}$ & $12\times
10^{-11}$ \\ \hline
$\phi_{21}$ & $\sim\pi/4$ & $\sim\pi/12-\pi/4$ & $\sim\pi/18-\pi/4$ \\ \hline
\end{tabular}
\end{center}
\caption{Baryon Asymmetry for the Case of Thermal Leptogenesis}
\label{tab:phase}
\end{table}

The constraint on $\phi_{21}$ is obtained from considerations of
Big-Bang nucleosynthesis, which requires \(3.7 \times 10^{-11}
\lesssim (Y_B)_{BBN} \lesssim 9 \times 10^{-11}\) \cite{expr1};
this is consistent with the CMB data \cite{expr2}, which
suggests somewhat higher values of \((Y_{B})_{CMB} \approx
(7-10)\times 10^{-11}\) (say). We see that the first case
\(|\zeta_{31}-z| \approx 1/200(1/3)\) leads to a baryon asymmetry
$Y_B$ that is on the low side of the BBN-data, even for a maximal
\(\sin(2\phi_{21})\approx 1\). The other cases with
$|\zeta_{31}-z|\approx (1/200)(1/4\mbox{ to }1/5)$, which
 are of course
perfectly plausible, lead to the desired magnitude of the baryon asymmetry
for natural values of the phase parameter
\(\phi_{21}\sim (\pi/18\mbox{ to }\pi/4)\). We see that, for the thermal case,
the CMB data, requiring higher values of $Y_{B}$,
 would suggest somewhat smaller values of
\(|\zeta_{31} -z| \sim 10^{-3}\).
This constraint would be eliminated for the case of non-thermal leptogenesis.

We next consider briefly the scenario of non-thermal leptogenesis
\cite{41, Kumekawa}. In this case the
inflaton is assumed to decay, following the inflationary epoch, 
directly into a pair of heavy RH neutrinos (or sneutrinos).
These in turn decay into $l+H$ and $\bar{l} +\bar{H}$ as well as
into the corresponding SUSY modes, and thereby produce lepton asymmetry, during
the process of reheating. It turns out that this scenario
goes well with the fermion mass-pattern of Sec. 2 [in particular see Eq.
\eqref{eq:MajM}] and the observed baryon asymmetry, provided
$2M_2>m_{\rm infl}>2M_1$, so that the inflaton
decays into $2N_1$ rather than into $2N_2$ (contrast this from the case
proposed in Ref.~\cite{41}). In this case, the reheating
temperature ($T_{\rm RH}$) is found to be much less than
$M_1\sim 10^{10}$ GeV (see below); thereby
(a) the gravitino constraint is satisfied quite easily, even for a rather
low gravitino-mass $\sim 200$ GeV (unlike in the thermal case); at the same
time (b) while $N_1$'s are produced non-thermally (and copiously) through
inflaton decay, they remain out of equilibrium and the wash out process
involving inverse decays and $\Delta L\neq 0$-scatterings are ineffective,
so that the efficiency factor $\kappa$ is 1.

To see how the non-thermal case can arise naturally, we recall
that the VEV's of the Higgs fields $\Phi=(1,2,4)_H$ and $\bar
\Phi=(1,2,\bar 4)_H$ have been utilized to (i) break SU(2)$_R$ and
B-L so that G(224) breaks to the SM symmetry~\cite{PS}, and
simultaneously (ii) to give Majorana masses to the RH neutrinos
via the coupling in Eq.~\eqref{eq:LMaj} (see e.g., Ref.~\cite{BPW}); 
for SO(10), $\bar \Phi$ and $\Phi$ would be in ${\bf
16}_H$ and $\bar{\bf 16}_H$ respectively), It is attractive to
assume that the same $\Phi$ and $\bar\Phi$ (in fact their
$\tilde{\nu_{\rm RH}}$ and $\bar{\tilde{\nu}}_{\rm
RH}$-components), which acquire GUT-scale VEV's, also drive
inflation \cite{41}. In this case the inflaton
would naturally couple to a pair of RH neutrinos by the coupling
of Eq.~\eqref{eq:LMaj}. To implement hybrid inflation in this
context, let us assume following Ref.~\cite{41},
an effective superpotential \(W_{\rm eff}^{\rm infl}=\lambda
S(\bar\Phi\Phi-M^2)+\mbox{(non-ren. terms)}\), where $S$ is a
singlet field \cite{newFN55}. It has been shown in 
Ref.~\cite{41} that in this case a flat potential with
a radiatively generated slope can arise so as to implement
inflation, with $G(224)$ broken during the inflationary epoch to
the SM symmetry. The inflaton is made of two complex scalar fields
(i.e., \(\theta=(\delta\tilde\nu_H^C+\delta\tilde{\bar \nu}_H^C)/
\sqrt{2}\) that represents the fluctuations of the Higgs fields
around the SUSY minimum, and the singlet S). Each of these have a
mass $m_{\rm infl}=\sqrt{2}\lambda M$, where \(M=\langle\mbox{(1,
2, 4)}_H\rangle\approx 2\times 10^{16}\) GeV and a width
\(\Gamma_{\rm
infl}=\Gamma(\theta\rightarrow\Psi_{\nu_H}\Psi_{\nu_H}) =
\Gamma(S\rightarrow\tilde\nu_H\tilde\nu_H) \approx
[1/(8\pi)](M_1/M)^2{m}_{\rm infl}\) so that
\begin{equation}
T_{\rm RH}\approx (1/7)(\Gamma_{\rm infl}M_{\rm Pl})^{1/2} \approx (1/7)(M_1/M)
[m_{\rm infl}M_{\rm Pl}/(8\pi)]^{1/2}
\end{equation}
For concreteness, take \cite{newFN56}
\(M_2\approx 2\times 10^{12}\) GeV, \(M_1\approx 2\times 10^{10}\) GeV (1 to 2)
[in accord with Eq. \eqref{eq:MajM}], and $\lambda\approx 10^{-4}$, so
that ${m}_{\rm infl}\approx 3\times 10^{12}$ GeV.
We then get:
\(T_{\rm RH}\approx (1.7\times 10^8\mbox{ GeV})\mbox{(1 to 2)}\),
and thus (see e.g., Sec. 8 of
Ref.~\cite{d}):
\begin{eqnarray}\label{YB2}
(Y_B)_{Non-Thermal}&\approx& -(Y_L/3) \nonumber\\
&\approx& (-1/3) [(n_{N_1}+n_{\tilde N_1}+n_{\bar{\tilde{N}}_1})/s]\epsilon_1
\nonumber\\
&\approx& (-1/3)[(3/2)(T_{\rm RH}/m_{\rm infl})\epsilon_1] \nonumber\\
&\approx& (30\times 10^{-11})(\sin 2\phi_{21})\mbox{(1 to 2)}^2
\end{eqnarray}
Here we have used Eq.~\eqref{eq:epsilon2} for $\epsilon_1$ with
appropriate $(M_1/M_2)$, as above. Setting $M_1\approx 2\times
10^{10}$ for concreteness, we see that $Y_B$ obtained above agrees
with the (nearly central) observed value of \(\langle
Y_B\rangle^{\rm central}_{\rm BBN(CMB)} \approx (6(9)) \times
10^{-11}\), again for a natural value of the phase parameter
$\phi_{21}\approx \pi/30 (\pi/20)$. As mentioned above, one
possible advantage of the non-thermal over the thermal case is
that the gravitino-constraint can be met rather easily, in the
case of the former (because $T_{RH}$ is rather low
$\sim 10^{8}$ GeV), whereas for the thermal case there is a
significant constraint on the lowering of the $T_{RH}$ (so as to
satisfy the gravitino-constraint) vis a vis a raising of $M_1 \sim
T_{RH}$ so as to have sufficient baryon asymmetry (note that
$\epsilon_1 \propto M_1$, see Eq.~(\ref{eq:epsilon2})).
Furthermore, for the non-thermal case, the dependence of $Y_B$ on
the parameter $|\zeta_{31} -z|^2$ (which arises through $\kappa$
and $\tilde{m}_1$ in the thermal case, see Eqs. (\ref{eq:d}),
(\ref{eq:k}), and (\ref{eq:terms11})) is largely eliminated. Thus
the expected magnitude of $Y_{B}$ (Eq.~(\ref{YB2})) holds without
a significant constraint on $|\zeta_{31} - z|$ (in contrast to the
thermal case).

To conclude, we have considered two alternative scenarios (thermal as well as
non-thermal) for inflation and leptogenesis.
We see that the G(224)/SO(10) framework provides a simple and 
{\it unified description} of not only fermion masses and neutrino oscillations
(consistent with maximal atmospheric and large solar oscillation angles) but 
also of baryogenesis via leptogenesis, treated within either scenario, in accord with the gravitino-constraint. 
Each of the features -- (a) the existence of the right-handed neutrinos, (b)
B-L local symmetry, (c) quark-lepton unification through SU(4)-color, (d) the magnitude of the supersymmetric unification-scale and (e) the seesaw mechanism -- plays a crucial role in
realizing this unified and successful description. These features in turn
point to the relevance of either the G(224) or the SO(10)
symmetry being effective between the string and the GUT scales, in four 
dimensions \cite{FN1}.  While the observed magnitude of the baryon
asymmetry seems to emerge naturally from within the framework, understanding
its observed sign (and thus the relevant CP violating phases) remains a
challenging task \cite{effphase}.\\

{\bf \Large Acknowledgements}\\

I would like to thank Kaladi S. Babu for collaborative discussions on
CP violation, and Pasquale Di Bari and Qaisar Shafi for most helpful
correspondences and clarifications of their work. I have also benefitted
from discussions with Gustavo Branco and Tsutumo Yanagida on aspects of this
work. The
sabbatical support by the University of Maryland during the author's visit to
SLAC, as well as the hospitality 
of the Theory Group of SLAC, where this work was carried out, are gratefully
acknowledged. The work is supported in part by DOE grant no. 
DE-FG02-96ER-41015.

}

\begin{thebibliography}{29}
\bibitem{expr1} See e.g., K. A. Olive, G. Steigman and T. P. Walker, Phys. Rep.
{\bf 333}, 389 (2000). For a recent analysis based on Big Bang
nucleosynthesis (BBN), see B. D. Fields and S. Sarkar,
{\it ``Review of Particle Physics''}, Phys. Rev. {\bf D66}, 010001 (2002),
which yields: $Y_B^{BBN}\approx (\eta_B)^{BBN}/7\approx (3.7-9)\times
10^{-11}$.

\bibitem{expr2} For measurements of baryon asymmetry through observation of
acoustic peaks in the cosmic microwave background radiation, see
P. de Barnardis et al., Astrophys. J. {\bf 564}, 559 (2002)
(BOOMERanG experiment), and C. Pryke et al. Astrophys. J. {\bf 568}, 46
(2002) (DASI experiment). A combined analysis of these observations yield
[A. Benoit (the Archeops collaboration), astro-ph/0210306]:
$Y_B^{CMB}\approx (8.6\genfrac{}{}{0pt}{3}{+1.1}{-1.6})\times 10^{-11}$,
which is of course consistent with the BBN-value.
Most recently, the WMAP surveying the entire celestial sphere with high
resolution yields:
\((Y_{B})_{\mathrm{WMAP}} \approx (8.7 \pm 0.4) \times 10^{-11}\)
(WMAP collaboration, astro-ph/0302207--09--13--15,17,18,20,22--25).

\bibitem{Ruzmin} V. Kuzmin, V. Rubakov and M. Shaposhnikov, Phys. Lett. 
{\bf B155}, 36 (1985).

\bibitem{Yanagida} M. Fukugita and T. Yanagida Phys. Lett. 
{\bf B174}, 45 (1986); G. Lazarides and Q. Shafi, Phys. Lett, {\bf B258}, 305
(1991); M. A. Luty, Phys. Rev. {\bf D45}, 455 (1992). The second paper
discusses leptogenesis in the context of inflationary cosmology.

\bibitem{Others} For an incomplete list of references on further works on
leptogenesis see e.g.,
M. Flanz, E. A. Pascos and U. Sarkar, Phys. Lett. {\bf B345}, 248 (1995);
L. Covi, E. Roulet and F. Vissani, Phys. Lett. {\bf B384}, 169 (1996);
W. Buchm\"uller and M. Pl\"umacher, Phys. Lett. {\bf B389}, 73 (1996); ibid
{\bf B431}, 354 (1998); A. Pilaftsis, Phys. Rev. {\bf D56}, 5431 (1997);
T. Moroi and H. Murayama, hep-ph/9908223; M. S. Berger and B. Brahmachari,
Phys. Rev. {\bf D60}, 073009 (1999); J. Ellis, S. Lola and D. Nanopoulos,
Phys. Lett. {\bf B452}, 87 (1999); E. Ma, S. Sarkar and U. Sarkar, Phys. Lett.
{\bf B458}, 73 (1999); K. Kang, S. F. Kang and U. Sarkar, Phys. Lett. 
{\bf B486}, 391 (2000); R. Rangarajan and H. Mishra, Phys. Rev. {\bf D61},
043509 (2000); H. Goldberg, Phys. Lett. {\bf B474}, 389 (2000); R. Barbieri,
P. Creminelli, A. Strumia and N. Tetradis , Nucl. Phys. {\bf B575}, 61 (2000);
D. Falcone and F. Tramontano, hep-ph/0101151; H. B. Nielsen and Y. Takanishi,
Phys. Lett. {\bf B507}, 241 (2001); J. Ellis, M. Raidal and T. Yanagida,
ph/0206300; P. H. Frampton, S. L. Glashow and T. Yanagida, hep-ph/0208157;
M. Fujii and T. Yanagida, hep-ph/0208191; R.~Allahverdi and A.~Majumdar,
hep-ph/0208286; Zhi-Zhong Xing, hep-ph/0209066;
T. Endoh et al., hep-ph/0209098.
 For references to works on leptogenesis based on
SO(10)-models, see Ref. \cite{SO(10)}. 

\bibitem{GUT-baryo} To mention about a few alternative mechanisms, GUT-baryogenesis satisfying $\Delta (B-L)=0$ (as in minimal SU(5)) becomes ineffective because it is wiped out by electroweak sphaleron effects. Standard GUT-baryogenesis involving decays of X and Y gauge bosons (with $M_X \sim 10^{16}\mbox{ GeV}$) and/or superheavy Higgs bosons is hard to realize anyway within a plausible inflationary scenario satisfying the gravitino-constraint. [See e.g.\ E. W. Kolb and M. S. Turner, "The Early Universe", Addison-Wesley (1990)]. On the other hand, purely electroweak baryogenesis based on the sphaleron effects appears to be excluded for the case of the standard model without supersymmetry, and highly constrained as regards the available parameter space for the case of the supersymmetric standard model, owing to LEP lower limit on the Higgs mass $\geq 115\mbox{ GeV}$. For a recent review of these and other mechanisms and also for relevant references, see e.g.\ M. Dine and A. Kusenko, hep-ph/0303065.

\bibitem{PS} J. C. Pati and A. Salam, Phys. Rev. Lett. 
{\bf 31}, 661 (1973); Phys. Rev. {\bf D10}, 275 (1974).

\bibitem{Georgi} H. Georgi in 'Particles and Fields', Ed. by C. Carlson
(AIP, NY, 1975), 575; H. Fritzsch and P. Minkowski, Ann. Phys. {\bf 93}, 193 
(1975).

\bibitem{FN1} Promising string theory solutions yielding the G(224)-symmetry
in 4D have been obtain, using different approaches, by a number of authors;
see e.g., I. Antoniadis, G. Leontaris and J. Rizos, Phys. Lett. {\bf B245},
161, (1990); A. Murayama and A. Toon, Phys. Lett. {\bf B318}, 298 (1993);
G. Shiu and S. H. Tye, Phys. Rev. {\bf D58}, 106007 (1998); Z. Kakushadze,
Phys. Rev. {\bf D58}, 1010901 (1998); G. Aldazabal, L. E. Ibanez and
F. Quevedo, hep-th/9909172; C. Kokorelis, hep-th/0203187, hep-th/0209202; 
M. Cvetic, G. Shiu, and A. M. Uranga, \emph{Phys. Rev. Lett.} \textbf{87},
201801 (2001), hep-th/0107143, and \emph{Nucl. Phys.} \textbf{B615},
3 (2001), hep-th/0107166; M. Cvetic and I. Papadimitriou, hep-th/0303197;
R. Blumenhagen, L. Gorlich and T. Ott, 
hep-th/0211059. For a type I string-motivated scenario leading to the G(224)
 symmetry in 4D, see L. I. Everett, G. L. Kane, 
S. F. King, S. Rigolin and L. T. Wang, hep-th/0202100. Recently there have
also been several works based on orbifold compactifications of five and six dimensional GUT-theories which yield the G(224) symmetry in 4D with certain desirable features; See e.g. R. Dermisek and A. Mafi, Phys. Rev. {\bf D65}, 055002 (2002) [hep-ph/0108139]; Q. Shafi and Z. Tavartkiladze, hep-ph/0108247,
hep-ph/0303150; C. H. Albright and S. M. Barr, hep-ph/0209173; H. D. Kim and S. Raby, 
hep-ph/0212348; I. Gogoladze, Y. Mimura and S. Nandi, hep-ph/0302176; B. Kyae and Q. Shafi, hep-ph/0211059; H. Baer et al., hep-ph/0204108;
For a global phenomenological analysis of a realistic string-inspired 
supersymmetric model based on the $G(224)$ symmetry see T. Blazek,
S. F. King, and J. K. Perry (hep-ph/0303192); and also references therein. Possible advantages of a string-unified
four-dimensional G(224)-solution over an
SO(10)-solution, and vice-versa, are discussed in J. C. Pati, hep-ph/0204240
(see Sec. 3). The G(224)-solution on the one hand is free
from the problem of doublet-triplet splitting, and on the other hand can
account for the observed coupling unification, if the gap between $M_{\rm st}$
and $M_{\rm GUT}$ is small ($M_{\rm GUT}\approx 
(1/2\mbox{ to }1/3)M_{\rm st}$, say). Despite this small gap, one would,
however, 
have the benefits of SU(4)-color to understand fermion masses and neutrino
oscillations (as reviewed in Sec. 2) and implement baryogenesis via
leptogenesis, to be discussed here.

\bibitem{Gursey}F. Gursey, P. Ramond and R. Slansky, Phys. Lett, {\bf B60},
177 (1976).

\bibitem{GeorgiGlashow} H. Georgi and S. L. Glashow, Phys. Rev. Lett. {\bf 32},
438 (1974).

\bibitem{flipSU5} S. M. Barr, Phys. Lett. {\bf B112}, 219 (1982); J. P. Derendinger, 
J. E. Kim and D. V. Nanopoulos, Phys. Lett. {\bf B139}, 170 (1984); I. 
Antoniadis, J. Ellis, J. Hagelin and D. V. Nanopoulos, Phys. Lett. {\bf 
B194}, 231 (1987).

\bibitem{SU3} F. Gursey, P. Ramond and R. Slansky, Phys. Lett, {\bf B60}, 177
(1976); Y. Achiman and B. Stech, Phys. Lett. {\bf B77}, 389 (1978);
Q. Shafi, Phys. Lett. {\bf B79}, 301 (1978); A. deRujula, H. Georgi and 
S. L. Glashow, 5th Workshop on Grand Unification, edited by K. Kang et al.,
World Scientific, 1984, p88.

\bibitem{Pati_etal} J. C. Pati and A. Salam, Phys. Rev. {\bf D10}, 275
(1974); R. N. Mohapatra and J. C. Pati, Phys. Rev. {\bf D11}, 566 and 2558
(1975).

\bibitem{SUSYunifscale} S. Dimopoulos, S. Raby and F. Wilczek,
Phys. Rev. {\bf D 24}, 1681 (1981); W. Marciano and G. Senjanovic,
Phys. Rev. {\bf D 25}, 3092 (1982) and M. Einhorn and D. R. T. Jones,
Nucl. Phys. {\bf B 196}, 475 (1982).
For work in recent years, see P. Langacker and M. Luo,
Phys. Rev. {\bf D 44}, 817 (1991); U. Amaldi, W. de Boer and H. Furtenau,
Phys. Rev. Lett. {\bf B 260}, 131 (1991); F. Anselmo, L. Cifarelli,
A. Peterman and A. Zichichi, Nuov. Cim. {\bf A 104} 1817 (1991).

\bibitem{seesaw} M. Gell-Mann, P. Ramond and R. Slansky, in:  {\it
Supergravity}, eds. F. van Nieuwenhuizen and D. Freedman (Amsterdam,
North Holland, 1979) p. 315; T. Yanagida, in:  {\it Workshop on the
Unified Theory and Baryon Number in the Universe}, eds. O. Sawada and A.
Sugamoto (KEK, Tsukuba) 95 (1979); R. N. Mohapatra and G. Senjanovic,
Phys. Rev. Lett. {\bf 44}, 912 (1980).

\bibitem{SeeJCP} See e.g.,
J. C. Pati, ``{\it Implications of the SuperKamiokande
Result on the Nature of New Physics}'', in Neutrino 98, Takayama,
Japan, June 98, hep-ph/9807315; Nucl. Phys. B (Proc. Suppl.) {\bf 77},
Page 299 (1999).

\bibitem{Superk} SuperKamiokande Collaboration, Y. Fukuda et. al.,
Phys. Rev. Lett. {\bf 81}, 1562 (1998).

\bibitem{SeeReview} See e.g., Reviews in A. Pilaftsis, Int. J. Mod. Phys. 
{\bf A14}, 1811, (1999); W. Buchm\"uller and M. Pl\"umacher, Int. J. Mod. 
Phys. {\bf A15}, 5047 (2000).

\bibitem{BPW} K. S. Babu, J. C. Pati and F. Wilczek, hep-ph/9812538, Nucl.
Phys. {\bf 566}, 33 (2000).

\bibitem{JCP} J. C. Pati, {\it ``Confronting Conventional Ideas of Grand 
Unification
With Fermion Masses, Neutrino Oscillations and Proton Decay''},
Lectures given at the summer school, ICTP, Trieste (June, 2001), to appear in
the proceedings,
 hep-ph/0204240.

\bibitem{LMA} Q. R. Ahmad et al (SNO), Phys. Rev. Lett. {\bf 81}, 011301 (2002); B. T. Cleveland et al (Homestake),
Astrophys. J. {\bf 496}, 505 (1998); W. Hampel et al. (GALLEX), Phys. Lett. {\bf B447}, 127, (1999);
J. N. Abdurashitov et al (SAGE) (2000), astro-ph/0204245; M. Altmann eta al. (GNO),
Phys. Lett. {\bf B 490}, 16 (2000); S. Fukuda et al.
(SuperKamiokande), \emph{Phys. Lett.} \textbf{B539}, 179 (2002). 
Disappearance of $\bar{\nu}_e$'s produced in earth-based reactors
is established by the KamLAND data: K. Eguchi et al., hep-ex/0212021.

\bibitem{BabuJCPCascais} K. S. Babu and J. C. Pati, {\it ``Link Between 
Neutrino
Oscillations and CP Violation within Supersymmetric Unification''}, Abstract
in Proceedings of NATO Advanced Study Institute, ``Recent Developments in
Particle Physics and Cosmology'', June 6-July 7, 2000, Cascais, Portugal,
Pages 215-217, Ed. by G. Branco, Q. Shafi and J. I. Silva- Marcos; K. S. Babu
and J. C. Pati, {\it ``Linking CP and Flavor Violations To Neutrino 
Oscillations Within Supersymmetric Unification''} paper to appear.

\bibitem{FN2} By combining these results with the analysis of the forthcoming 
paper \cite{BabuJCPCascais}, one would incorporate CP violation into this
unified picture as well.

\bibitem{Incompletelist} Many of these are based on
phenomenological models just for neutrino masses, which are not linked to the
masses and mixings of quarks and charged leptons. See e.g., S. F. King,
hep-ph/0204360 for a recent analysis along these lines and references
therein.

\bibitem{SO(10)} For attempts within SO(10) models, see e.g., E. Nezri and 
J. Orloff, hep-ph/0004227; F. Bucella,
D. Falcone, F. Tramontano, hep-ph/0108172; G. C. Branco, F. Gonzales Felipe,
F. R. Joaquim and M. N. Rebelo, hep-ph/0202030. For a variant attempt within
left-right symmetric model, see A. Joshipura, E. A. Paschos and W. Rodejohann,
hep-ph/0104228.

\bibitem{FN3} For example, many of the attempts in \cite{SO(10)} assume that 
the Dirac
mass-matrix of the neutrinos is equal to that of the up-flavor quarks
($M_\nu^D=M_u$) at GUT-scale. This simple equality would be true for SO(10) if
only ${\bf 10}_H$ contributes to the fermion masses. However, the
minimal Higgs system permits a (B-L)-dependent antisymmetric "23" and "32"
entry \cite{BPW} (as discussed later), which plays a crucial role in
explaining why $m_\mu\neq m_s$ and why $V_{cb}$ is so small and yet
$\theta_{\nu_2\nu_3}^{\rm osc}$ is rather maximal. Such entries do not
respect $M_\nu^D=M_u$.

\bibitem{23} For instance, in the first paper of Ref. \cite{SO(10)}, it is 
found that only the solar vacuum oscillation solution gives acceptable
baryon asymmetry. In the second paper, it is noted that SUSY models with 
full quark-lepton symmetry gives too small an asymmetry, while in the third
paper it is found that the just-so and SMA solutions give viable leptogenesis,
but the LMA solution is strongly disfavored [based on their assumption of 
$M_\nu^D=M_u$, (see comments in Ref. \cite{FN3})]. In the fourth paper,
it is observed that the SMA and vacuum solutions produce reasonable asymmetry,
but the LMA solution produces too large an asymmetry.

\bibitem{Hall}   These have been introduced in various forms in the
literature. For a sample, see e.g., C. D. Frogatt and H. B. Nielsen, Nucl. 
Phys. {\bf B147}, 277 (1979); L. Hall and H. Murayama, Phys. Rev. Lett. 
{\bf 75}, 3985 (1995); P. Binetruy, S. Lavignac and P. Ramond, Nucl. Phys.
{\bf B477}, 353 (1996).
In the string theory context, see e.g., A. Faraggi, Phys. Lett.
{\bf B278}, 131 (1992).

\bibitem{FN4} The zeros in "11", "13" and "31" elements signify that they
are relatively small quantities (specified below). While the "22" elements
were set to zero in Ref. \cite{BPW}, because they are meant to be
$<$"23""32"/"33"$\sim 10^{-2}$ (see below), and thus unimportant for purposes
of Ref. \cite{BPW}, they are retained here, because such small
$\zeta_{22}^u$ and $\zeta_{22}^d$ [$\sim (1/3)\times 10^{-2}$ (say)] can
still be important for CP violation and thus leptogenesis.

\bibitem{FN26} For G(224), one can choose the corresponding sub-multiplets --
that is (1, 1, 15)$_H$, (1, 2, $\bar{4}$)$_H$, (1, 2, 4)$_H$, (2, 2, 1)$_H$
-- together with a singlet $S$, and write a superpotential analogous to Eq. 
\eqref{eq:Yuk}.

\bibitem{FN6} If the effective non-renormalizable operator like 
${\bf 16}_2{\bf 16}_3{\bf 10}_H{\bf 45}_H/M'$ is induced through exchange
of states with GUT-scale masses involving renormalizable couplings, rather
than through quantum gravity, $M'$ would, however, be of order GUT-scale. 
In this case $\langle {\bf 45}_H\rangle/M'\sim 1$, rather than 1/10.

\bibitem{FN7} While ${\bf 16}_H$ has a GUT-scale VEV along the SM singlet, it 
turns it cam also have a VEV of EW scale along the ``$\tilde\nu_L$''
direction due to its mixing  with ${\bf 10}_H^d$, so that the $H_d$ of 
MSSM is a mixture of  ${\bf 10}_H^d$ and  ${\bf 16}_H^d$.
This turns out to be the origin of non-trivial CKM mixings
 (See Ref. \cite{BPW}).

\bibitem{FN8}  The flavor charge(s) of ${\bf 45}_H$(${\bf 16}_H$) would
get determined depending upon whether $p$($q$) is one or zero (see below).

\bibitem{FN9} The basic presumption here is that effective dimensionless
couplings allowed by SO(10)/G(224) and flavor symmetries are of order unity
[i.e., $(h_{ij},g_{ij},a_{ij})\approx 1/3$-3 (say)]. The need for appropriate
powers of $(S/M)$ with $\langle S\rangle/M\sim M_{\rm GUT}/M_{\rm string}\sim
(1/10$-1/20) in the different couplings leads to a hierarchical structure. 
As an 
example, consider just one U(1)-flavor symmetry with one singlet S. The
hierarchical form of the Yukawa couplings exibited in Eqs.
\eqref{eq:mat} and  \eqref{eq:Yuk} would
be allowed, for the case of $p=1$, $q=0$, if (${\bf 16}_3$, ${\bf 16}_2$, 
${\bf 16}_1$,
${\bf 10}_H$, ${\bf 16}_H$, ${\bf 45}_H$ and S) are assigned U(1)-charges of
($a$, $a+1$, $a+2$, $-2a$, $-a-1/2$, 0, -1). 
It is assumed that other fields are present that would make the U(1) symmetry 
anomaly-free. With this assignment of charges, one
would expect $|\zeta_{22}^{u,d}|\sim (\langle S\rangle/M)^2$; one may thus take
$|\zeta_{22}^{u,d}|\sim (1/3)\times 10^{-2}$ without upsetting the success of
Ref. \cite{BPW}. In the same spirit, one would expect $|\zeta_{13},
\zeta_{31}|\sim (\langle S\rangle/M)^2\sim 10^{-2}$ and $|\zeta_{11}|\sim 
(\langle S\rangle/M)^4\sim 10^{-4}$ (say). The value of ``a'' would get fixed
by the presence of other operators (see later).

\bibitem{FN30} These effective non-renormalizable couplings can of course
arise through exchange of (for example) ${\bf 45}$ in the string tower,
involving
renormalizable ${\bf 16}_i{\bf \bar{16}}_H{\bf 45}$ couplings. In this case,
one would expect $M\sim M_{\rm string}$.

\bibitem{Kaplunovsky} P. Ginsparg, Phys. Lett. {\bf B197}, 139 (1987);
V. S. Kaplunovsky, Nucl. Phys. {\bf B307}, 145 (1988); Erratum: ibid
{\bf B382}, 436 (1992).

\bibitem{JCPErice} J. C. Pati, "{\it Probing Grand Unification Through Neutrino Oscillations, Leptogenesis and Proton Decay}", hep-ph/0305221. To appear in the Proceedings of the Erice 2002 School.

\bibitem{rangeM3} The range in $M_3$ and $M_2$ is constrained by the values of $m(\nu_3)$ and $m(\nu_2)$ suggested by the atmospheric and solar neutrino data.

\bibitem{newFN36} Note that the magnitudes of $\eta$, $\epsilon$ and
$\sigma$ are fixed by the input quark masses. Furthermore, one can argue
that the two contributions for $\theta_{\nu_2\nu_3}^{\rm osc}$
[see Eq. \eqref{eq:pred}] necessarily add to each other as long as $|y|$ is
hierarchical ($\sim 1/10$) \cite{BPW}. As a result, once the sign of $\epsilon$
relative to $\eta$ and $\sigma$ is chosen to be negative, the actual
magnitudes of $V_{cb}\approx (0.044)$ and $\sin^2
2\theta_{\nu_2\nu_3}^{\rm osc}\approx
0.99$ emerge as predictions of the model \cite{BPW}.

\bibitem{FN32} Note that such an operator would be allowed by the flavor
symmetry defined in Ref. \cite{FN9} if one sets $a=1/2$. In this case,
operators such as $W_{23}$ and $W_{33}$ that would contribute to
$\nu_L^\mu\nu^\tau_L$ and $\nu_L^\tau\nu^\tau_L$ masses would be suppressed
relative to $W_{12}$ by flavor symmetry. As pointed out by other authors
(see e.g., S. Weinberg, Phys. Rev. Lett. {\bf 43}, 1566 (1979) and Proc. 
XXVI Int'l Conf. on High Energy Physics, Dallas, TX, 1992; E. Akhmedov, Z.
Berezhiani and G. Senjanovic, Phys. Rev. {\bf D47}, 3245 (1993).), 
non-seesaw Majorana masses of the LH neutrinos 
can arise directly, even in the standard model, through operators of the form
$L_iL_j\Phi_H\Phi_H/M$, by utilizing quantum gravity. [For SO(10), two 
${\bf 16}_H$'s are needed additionally to violate B-L by two units.] In the
case of the standard model, ordinarily, one would expect 
$M\sim M_{\rm Planck}$. Thus one would still
need to find a reason (in the context of the standard model) why (a)
$M\sim M_{\rm GUT}$ and also (b) why $L_1L_2\Phi_H\Phi_H/M$ is the leading 
operator in its class, rather than being suppressed (due to flavor symmetries)
relative to $L_3L_3\Phi_H\Phi_H/M$ (for example). Both (a) and (b) are needed
for this direct non-seesaw mass to be relevant to the LMA MSW solution.

\bibitem{FN33} A term like $W_{12}$ can be induced in the presence of, for 
example, a singlet $\hat S$ and a ten-plet (${\bf\hat{10}}$), possessing 
effective renormalizable couplings of the form $a_i{\bf 16}_i{\bf 16}_H
{\bf\hat{10}}$, $b{\bf\hat{10}}{\bf 10}_H\hat S$ and mass terms $\hat M_S\hat S
\hat S$ and $\hat M_{10}{\bf\hat{10}}{\bf\hat{10}}$. In this case 
$\kappa_{12}/M_{\rm eff}^3\approx a_1a_2b^2/(\hat M_{10}^2\hat M_S)$. Setting
the charge $a=1/2$ (see Ref. \cite{FN9} and \cite{FN32}), and assigning charges
(-3/2, 5/2) to $({\bf\hat{10}}, \hat S)$, the couplings $a_1$, and $b$ would 
be
flavor-symmetry allowed, while $a_2$ would be suppressed but so also would be
the mass of ${\bf\hat{10}}$ compared to the GUT-scale. One can imagine that 
$\hat S$ on the other hand acquires a GUT-scale mass through for example the 
Dine-Seiberg-Witten mechanism, violating the U(1)-flavor symmetry. One can
verify that in such a picture, one would obtain $\kappa_{12}/M_{\rm eff}^3
\sim 1/M_{\rm GUT}^3$.

\bibitem{FN34} For instance, consider the superpotential for ${\bf 45}_H$ only:
$W({\bf 45}_H)=M_{45}{\bf 45}_H^2+\lambda {\bf 45}_H^4/M$, which yields
(setting $F_{{\bf 45}_H}=0$), either $\langle{\bf 45}_H\rangle=0$, or 
$\langle{\bf 45}_H\rangle^2=-[2M_{45}M/\lambda]$. Assuming that
``other physics'' would favor $\langle{\bf 45}_H\rangle\neq 0$, we see that
$\langle{\bf 45}_H\rangle$ would be pure imaginary, if the square bracket is
positive, with all parameters being real. In a coupled system, it is
conceivable that $\langle{\bf 45}_H\rangle$ in turn would induce phases (other
than "0" and $\pi$) in some of the other VEV's as well, and may itself become
complex rather than pure imaginary.

\bibitem{Wolfenstein} L. Wolfenstein, Phys. Rev. Lett. {\bf 51}, 1945 (1983).

\bibitem{FN36} Within the framework developed in Ref. \cite{BabuJCPCascais}, 
the CP violating phases entering into the SUSY contributions (for example
those entering into the squark-mixings) also arise entirely through phases
in the fermion mass matrices.

\bibitem{FN37} An intriguing feature is the prominence of the 
$\delta_{RR}^{23}(\tilde b_R\rightarrow \tilde s_R)$-parameter which
gets enhanced in part because of the largeness of the $\nu_2$-$\nu_3$
oscillation angle. This leads to large departures from the
predictions of the standard model, especially in transitions such as
$B_s\rightarrow \bar B_s$ and 
$B_d\rightarrow\Phi K_s\,(b\rightarrow s\bar s s)$ \cite{BabuJCPCascais}. This
feature has independently been noted recently by D. Chang, A. Massiero, 
and H. Murayama (hep-ph/0205111).

\bibitem{FN38} As an example, one such fit with complex parameters assigns
\cite{BabuJCPCascais}: $\sigma=0.10-0.012\,i$, $\eta=0.12-0.05\,i$,
$\epsilon=-0.095$, $\eta'=4.0\times 10^{-3}$, $\epsilon'=1.54\times 10^{-4}
e^{i\pi/4}$, $\zeta_{22}^u=1.25\times 10^{-3}e^{i\pi/9}$ and
$\zeta_{22}^d=4\times 10^{-3}e^{i\pi/2}$, ${\cal M}_u^0\approx 110$ GeV,
${\cal M}_D^0\approx 1.5$ GeV, $y\approx -1/17$ (compare with Eq. 
\eqref{eq:fit}
for which $\zeta_{22}^u=\zeta_{22}^d=0$). One obtains as outputs:
$m_{b,s,d}\approx(\mbox{5 GeV, 132 MeV, 8 MeV})$, $m_{c,u}\approx
(\mbox{1.2 GeV, 4.9 MeV})$, $m_{\mu, e}\approx(\mbox{102 MeV, 0.4 MeV})$
with $m_{t,\tau}\approx(\mbox{167 GeV, 1.777 GeV})$, $(V_{us}, V_{cb},
|V_{ub}|, |V_{td}|)\approx(0.217, 0.044, 0.0029, 0.011)$, while preserving
the predictions for neutrino masses and oscillations as in
Eq. \eqref{eq:pred}. The above serves to demonstrate that complexification 
of parameters of the sort presented above can
preserve the successes of Eq. \eqref{eq:pred} (\cite{BPW}). This particular
case leads to
$\eta_W=0.29$ and $\rho_W=-0.187$ \cite{BabuJCPCascais}, to be compared with
the corresponding standard model values (obtained from $\epsilon_K$, $V_{ub}$ 
and $\Delta m_{Bd}$) of $(\eta_W)_{\rm SM}\approx 0.33$ and $(\rho_W)_{\rm SM}
\approx +0.2$. The consistency of
such values for $\eta_W$ and $\rho_W$ (especially reversal of the sign of
$\rho_W$ compared to the SM value), in the light of having both standard
model and SUSY-contributions to CP and flavor-violations, and their
distinguishing tests, are discussed in Ref. \cite{BabuJCPCascais}.

\bibitem{d} For reviews, see chapters 6 and 8 in E. W. Kolb and M. S. Turner, 
``The Early Universe'', Addison-Wesley, 1990.

\bibitem{gravitino} J. Ellis, J. E. Kim and D. Nanopoulos, Phys. Lett.
{\bf 145B}, 181 (1984); M. Yu. Khlopov and A. Linde, Phys. Lett. {\bf 138B},
265 (1984); E. Holtmann, M. Kawasaki, K. Kohri and T. Moroi, hep-ph/9805405.

\bibitem{Campbell} B. A. Campbell, S. Davidson and K. A. Olive, Nucl. Phys.
{\bf B399}, 111 (1993).

\bibitem{Covi} L. Covi, E. Roulet and F. Vissani, Phys. Lett. {\bf B384},
169 (1996).

\bibitem{Plumacher} M. Plumacher, hep-ph/9704231.

\bibitem{Bari} P. Di Bari, hep-ph/0211175; W. Buchmuller, P. Di Bari and
M. Plumacher, Nucl. Phys. {\bf B643}, 367 (2002) [hep-ph/0205349].

\bibitem{newFN50} The factor 0.7 in Eq. \eqref{eq:d} [instead of 1 in Eq.
(14) of Ref.~\cite{Bari}] is an estimate that incorporates the modification
needed for SUSY corresponding to a doubling of $N_1$-decay width owing to
the presence of both $l+H$ and $\tilde l+\tilde H$-modes and an increase
of $g^*$ from 106 for the standard model to 228 for SUSY.

\bibitem{newFN51} One can verify that $K\equiv (\Gamma(N_1)/2H)_{T=M_1}
\approx (0.37)[M_{\rm Pl}/(1.66 \sqrt{g^*}(8\pi v^2))]\tilde{m_1}\approx
234 (\tilde{m_1}/{\rm eV})$, where 0.37 denotes the usual time-dilation
factor, $g^*{\rm (for~SUSY)}\approx 228$ and $v\approx 174$ GeV. For
comparison, we note that if one includes only inverse decays (thus
neglecting $\Delta L\neq 0$-scatterings) in the Boltzmann equations, one
would obtain: $\kappa\approx 0.3/[K(\ln K)^{0.6}]$ for $K>10$ \cite{d}, and
$\kappa\approx 1/2K$ for $1\lesssim K\lesssim 10$. As pointed out in Ref.
\cite{Bari}, these expressions, frequently used in the literature, however,
tend to overestimate $\kappa$ by nearly a factor of 7. In what follows, we
will therefore use Eq. \eqref{eq:d} to evaluate $\kappa$.

\bibitem{m_tilde}M. Plumacher, Z. Phys. {\bf C74}, 549 (1997).


\bibitem{41} For a specific scenario of inflation and leptogenesis
in the context of SUSY G(224),
see R. Jeannerot, S. Khalil, G. Lazarides and Q. Shafi, JHEP {\bf 010},
012 (2000) (hep-ph/0002151), and references therein.
As noted in this paper, with the VEV's of
(1, 2, 4)$_H$ and (1, 2, $\bar 4$)$_H$ breaking G(224) to the standard model,
and also driving inflation, just the COBE measurement of 
$\delta T/T\approx 6.6\times 10^{-6}$, interestingly enough, implies that the
relevant VEV should be of order $10^{16}$ GeV. 
In this case, the inflaton made of two complex scalar fields (i.e.,
$\theta=(\delta\tilde\nu_H^c+\delta\tilde{\bar\nu}_H^c)/\sqrt{2}$, given by 
the fluctuations of the Higgs fields, and a singlet $S$), each 
 with a mass $\sim 10^{12}$-$10^{13}$ GeV, would decay {\it directly} into
a pair of heavy RH neutrinos -- that is into $N_2N_2$ (or  $N_1N_1$) if
$m_{\rm infl}>2M_2$ (or $2M_1$).
The subsequent decays of $N_2$'s (or $N_1$'s), thus produced, 
into $l+\Phi_H$ and 
$\bar l+\bar\Phi_H$ would produce lepton-asymmetry {\it during the process of
reheating}.
I will comment later on the consistency of this possibility with the fermion
mass-pattern exhibited in Sec. 2. 
I would like to thank Qaisar
Shafi for a discussion on these issues.

\bibitem{Kumekawa} K. Kumekawa, T. Moroi and T. Yanagida, Prog. Theor. Phys.
{\bf 92}, 437 (1994); G. F. Giudice, M. Peleso, A. Riotto and T. Tkachev,
JHEP {\bf 9908}, 014 (1999) [hep-ph/9905242]; T. Asaka, K. Hamaguchi,
M. Kawasaki and T. Yanagida, Phys. Lett. {\bf B464}, 12 (1999)
[hep-ph/9906366].

\bibitem{newFN55} Incorporating such an effective superpotential in accord
with the assignment of flavor-changes suggested in Refs. \cite{FN9}
and \cite{FN33} would involve two additional singlets with appropriate
charges. The (VEV)$^2$ of one or both of these may represent $M^2$.
Derivation of such a picture with appropriate flavor-charge assignments
from an underlying (string/M) theory is of course beyond the state of the
art at present.

\bibitem{newFN56} Note that for this non-thermal case, since the
gravitino-constraint is relaxed, $N_1$ can be chosen heavier than for the
case considered before (the thermal case), still in accord with Eq.
\eqref{eq:MajM}. Since $Y_B\propto\epsilon_1 T_{\rm RH}/{\cal M_{\rm infl}}$,
while $\epsilon_1\propto (M_1/M_2)$, $T_{\rm RH}\propto M_1
({\cal M_{\rm infl}})^{1/2}$ and ${\cal M_{\rm infl}}\propto\lambda$, we see
that $Y_B\propto (M_1^2/M_2)/\sqrt{\lambda}$, for a constant $M$, for the
case of non-thermal leptogenesis.

\bibitem{effphase} Note that the effective phase $\phi_{21}$, relevant to 
leptogenesis, depends on the phases in both the Dirac ($M_\nu^D$) 
and the Majorana ($M_R^\nu$) mass matrices of the neutrinos. Thus, 
in general, it is quite distinct from the phase(s) entering into 
observed CP violations in the K and the B-systems.

%\bibitem{epsilon1} L. Covi, E. Roulet and F. Vissani, in Ref. \cite{Others}.

%\bibitem{variant} A variant expression $[d=1/(2\sqrt{k^2+9})]$ for the dilution
%factor has also been used in the literature for lower values of $k$
%$(0\leq k\leq 10)$ [see e.g., H. B. Nielsen and Y. Takanishi, Phys. Lett.
%{\bf B507}, 241 (2001)]. This would yield a value for $d$ about (2 to 6)
%times lower than that obtained from Eq. \eqref{eq:d}, for
%$k\approx (2$ to 0.5). This should be kept in mind in viewing the results
%especially in the last column of Table 1.

%\bibitem{new45} Because of supersymmetry, lepton asymmetry should of course
%receive contributions from out-of-equilibrium decays of heavy sneutrinos
%($\tilde N_1$'s), and one must include the ``light'' sleptons in the decay
%and scattering processes, as well. These have not been included in our
%considerations for the sake of simplicity. We do not, however, expect them
%to alter the lepton asymmetry obtained as above by more than a factor of two.

%\bibitem{FN45} Note that for this scenario (with the inflaton decaying into
%$2N_1$) the gravitino constraint is very well satisfied even for
%$m_{3/2}\sim 300$ GeV, because the reheating temperature is rather low
%($\sim 10^8$ GeV). At the same time, one can allow $N_1$ to be heavier
%(like $2\times 10^{10}$ GeV) than the case considered before
%(like $4\times 10^{9}$ GeV) because it can be produced directly by the decay
%of the inflaton rather than from the thermal bath. Since $Y_B\propto
%\epsilon_1 T_{RH}$, $Y_B$ increases as $M_1^2$ for a given $M_2$. Thus, somewhat
%higher values of $M_1$ compatible with the range shown in Eq. \eqref{eq:MajM},
%are prefered.

\end{thebibliography}
\end{document}